\documentclass{aa}  
\usepackage{graphicx}
\usepackage{txfonts}
\usepackage{longtable}
\usepackage{array}

\usepackage{natbib}
\bibpunct{(}{)}{;}{a}{}{,} 

\begin{document}
   \title{Cool dwarfs stars from the Torino Observatory Parallax Program}

   \author{R.L. Smart\inst{1}
          \and
          G. Ioannidis\inst{2}
          \and
          H.R.A. Jones\inst{2}
          \and
          B. Bucciarelli\inst{1}
          \and
          M.G. Lattanzi\inst{1}}

   \offprints{R.L. Smart, smart@oato.inaf.it}

   \institute{
   INAF/Osservat\'orio Astronomico di Torino,
              Strada Osservat\'orio 20, 10025 Pino Torinese, Italy
	 \email{smart@oato.inaf.it}
          \and
   Centre for Astrophysics Research, Science and Technology Research Institute,
             University of Hertfordshire, Hatfield AL10 9AB
             }

   \date{Received 09/10/2009, accepted 02/03/2010}
 
     \abstract
     {}
     {We investigate and parameterise high proper motion red stars in the
       Torino Observatory Parallax Program. }
     {Observations of 27 objects were made over the period 1994 - 2001 on the
       1.05m Torino telescope.  The trigonometric parallaxes and proper
       motions were determined using standard techniques.}
     {We determine parallaxes and proper motions, and by comparison to models
       we infer masses, ages, and {metallicities}. Of the 27 objects, 22 are
       within 25pc and 4 appear to be subdwarfs. There are published
       parallaxes for 18 objects, and all but 4 agree to within 2$\sigma$. The
       discrepancies are discussed. }
     {}

\titlerunning{TOPP Cool dwarfs}
   \maketitle

\section{Introduction}

Stars of spectral type M make up over 70\% of the stars and 40\% of the mass
in our Galaxy.  Their main sequence lifetimes can exceed the age of the
universe, and they have age indicators that can be calibrated
\citep*[e.g.][]{2002AJ....123.3356G} to make them useful chronometers. A wide
variety of problems in Astronomy from the search for earth-like exoplanets
\citep{Torres2007} to modeling the Galaxy \citep{Juri'c2008} are addressed
using observations of these objects.

\begin{figure}
  \centering
  \includegraphics[width=7cm]{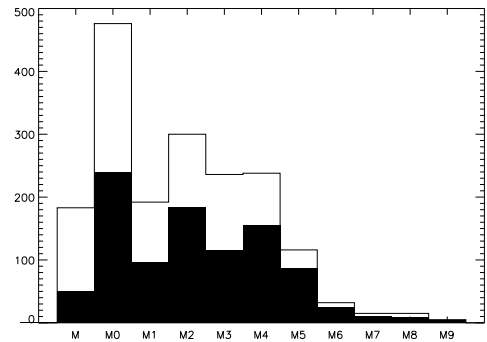}
  \caption{M dwarfs with parallaxes as a function of subclass from a
    combination of the Hipparcos 
    Yale Trigonometric Parallax and RECONS catalogues. The filled-in area
    represents those objects with errors of less than 10\%.}
  \label{mdwarfs}%
\end{figure}

Notwithstanding their numerical dominance and potential scientific use, M stars
make a surprising small proportion of objects that have a measured parallax.  In
the combined Hipparcos
\citep{Perryman1997} 
Yale Trigonometric Parallax \citep{2001yCat.1238....0V} and RECONS
\citep[Research Consortium on Nearby Stars, ][]{Henry2004} catalogues there are
over 120,000 stars of which less than 2000 are M dwarfs.  In Figure
\ref{mdwarfs} we plot the number of M dwarfs with parallaxes {as a function} of
spectral class. Approximately one half have relative parallax errors of less
than 10\% as shown by the shaded area. For the types later than M5 the total
number per bin is always less than 30. This is primarily because they are
intrinsically low-luminosity objects; of the 40,000 stars with apparent
magnitude V $<$ 8.0 in the Hipparcos catalogue there are only four M dwarfs.

The majority of the late M dwarfs that have measured parallaxes have been
determined using CCD observations, e.g. USNO \citep{Monet1992} and RECONS
programs. Here we present the M dwarfs found in the Torino Observatory
Parallax Program \citep[hereafter TOPP]{Smart2003} and we use recent models to
estimate mass and age for these objects. In section 2 we present the TOPP
observations and reduction procedures, in section 3 we present the resulting
parallaxes and proper motions, in section 4 we determine mass and age, and in
section 5 we discuss individual objects.

\section{Observations and Reduction Procedures }

\begin{figure}
  \centering
  \includegraphics[width=8cm]{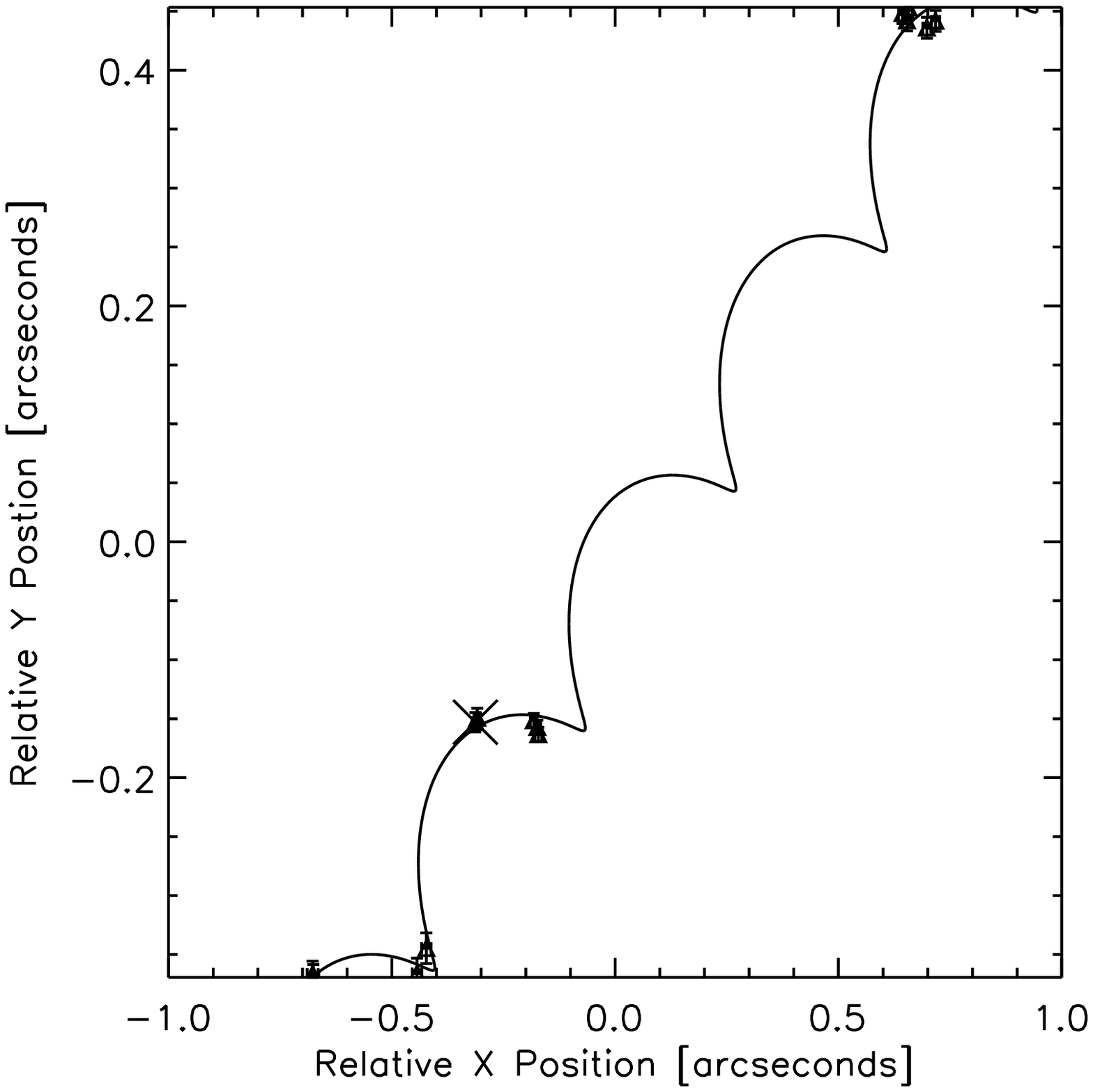} 
  \includegraphics[width=8cm]{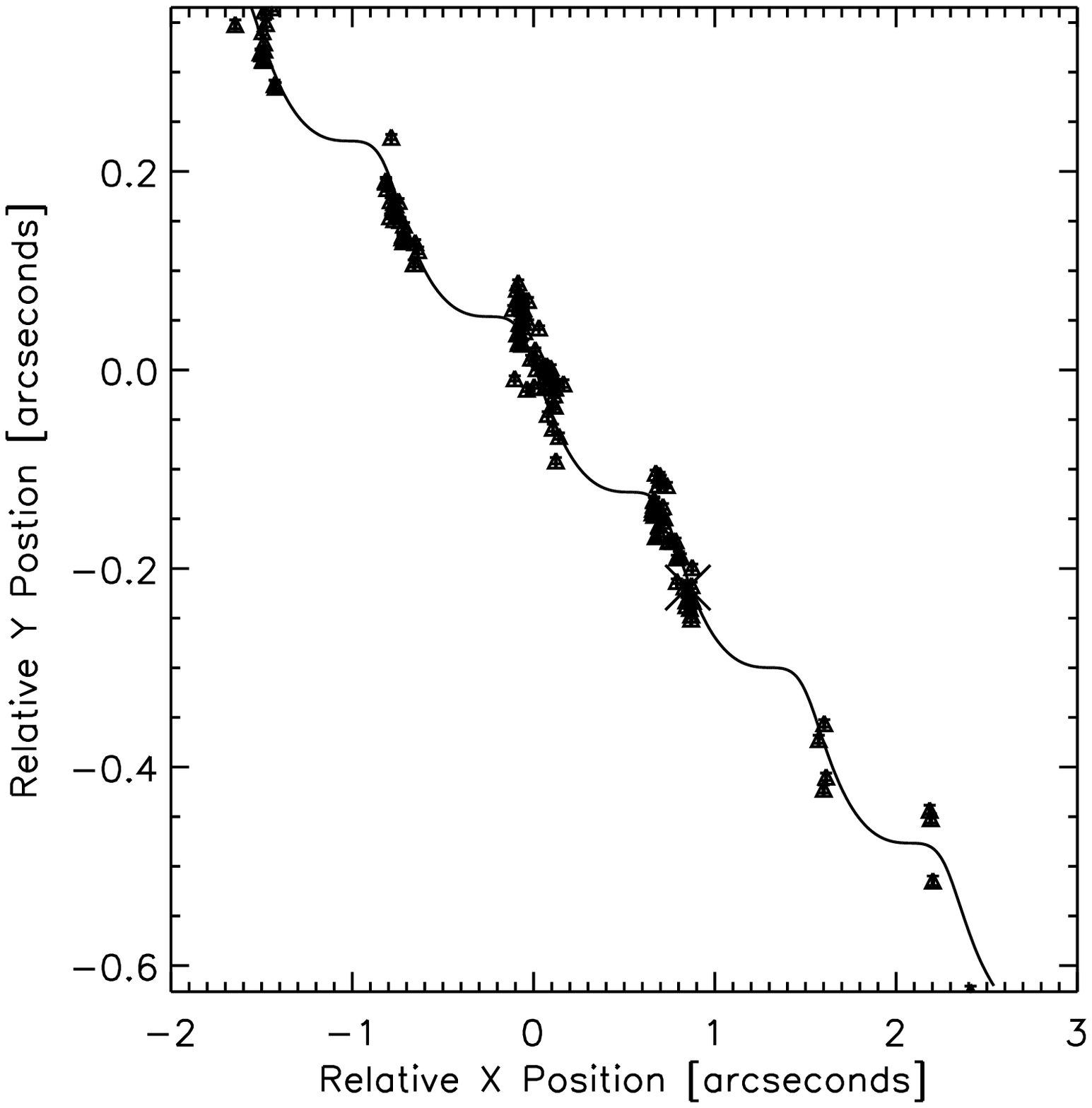} 
  \caption{Observations for two targets showing the range of coverage, 14
    observations for GJ 1167 A and 136 for LHS 1104.}
  \label{twosolutions}%
\end{figure}

The procedures for observation, image treatment and parallax determination
follow those described in previous papers \citep{Smart1999,Smart2003}. These
objects are very red compared to the field stars so a correction for
differential colour refraction as presented in \citet{Smart2007} is
applied. Here we outline the major steps of data reduction and the reader is
referred to those publications for further details.

Observations were all carried out on the Torino Observatory 1.05m reflecting
telescope which is a scaled-down version {of the 1.55m Kaj Strand Astrometric
Reflector at the USNO Flagstaff Station \citep{1966VA......8....9S}.} The CCD
used was an EEV CCD05-30 1296x1152 @ 15 microns/pixel constructed by the
Astromed company which provided a pixel scale of 0.47" and a field of view of
10'x9'. All parallax observations were carried out in the Cousins I filter.

The images are flat-fielded using sky flats taken each night and
bias-corrected using the overscan region of the CCD. All objects are found on
each frame and centered using the Robin software \citep{Lanteri1990}. A base
frame is chosen with the criteria that it was made in good seeing and in
the middle of the parallax observational sequence. Anonymous reference stars
are selected automatically following two criteria: they are on at
least 80\% of the frames, the residual 
between predicted and observed coordinates is less than three times the
overall mean residual. Selection of frames is also carried out automatically
eliminating any frames that{ have } a small number of stars in common with
the base frame ($<$ 9); and/or positional residuals larger than
three times the average frame residual.

A given sequence is iterated to obtain proper motions and parallaxes for all
objects and the above criteria are applied until the parallax of the target
changes by less than 1\%.  From the target star's relative parallax we must
subtract the mean parallax of the reference{ stars as we  }have implicitly
assumed that the reference frame is at a mean parallax of zero.  Using the
Mendez \& van Altena (1996\nocite{Mendez1996}) galaxy model we can calculate
the most probable distance of each reference star based on their
magnitude. The mean of these distances is an estimate of the correction to
absolute parallax that we add to the target's relative parallax (COR, in
{Table} \ref{parallax}). Relative parallax errors are found from the formal
scatter about the final fit; to this we add, in quadrature, 30\% of the
relative to absolute correction which we estimate to be the error of the
galactic model procedure used \citep{Smart1997}.

We note that all these targets were solved assuming a simple single star
solution; an extension to double star solutions will be made for those stars
with suspect or visual companions. For the visual binaries in this study the
orbital period is of the order of centuries so the difference will be
negligible. For the non-resolved binaries we still need to develop appropriate
software and routines; any effect, if present, will be reflected in the errors
(e.g. LHS 1976 discussed in Section 5).

\section{Results}

\begin{table*}
\caption{\label{parallax}Parallaxes and proper motions for TOPP red dwarfs.}
\centering
\begin{tabular}{llrrrrrrrrrr}
\hline\hline
ID, LHS &  GJ   & RA ~~~     & Dec ~~~     &$N_*,N_f$& $\mu_{\alpha}$      & $\mu_{\delta}$      &$\Delta$T&  COR  &  $\pi$        & Literature $\pi$\\
        &       & (h:m:s ~~)   & ($^\circ$:':'') ~~   &         &   (mas/yr)           &   (mas/yr)          & (yrs)      &  (mas)  &      (mas)     &    (mas)  \\
\hline  \\
 1,1047 & 1005  &  0:15:28.2 &$ -16$: 7:43.5 &   7, 37 &   596.8 $\pm$  4.7 &  -629.0 $\pm$  5.1 &   3.29 &   2.05 & 179.8 $\pm$ 11.6 & 191.9 $\pm$ 17.2$^1$\\
 2,1050 & 12    &  0:15:52.3  &$ +13$:32:41.0 &   8, 83 &   606.7 $\pm$  1.4 &   336.2 $\pm$  1.5 &   5.20 &   1.86 &  87.4 $\pm$  3.4 &    84 $\pm$ 11$^2$    \\
 3,1104 & ...   &  0:35:47.9  &$ +52$:41:41.7 &  33,136 &   775.9 $\pm$  1.0 &  -174.3 $\pm$  1.5 &   5.20 &   1.05 &  40.5 $\pm$  2.2 & ~~~~~~~ ... ~~~~~~~~~          \\
 4,1475 & 119 A &  2:56:29.6 &$ +55$:26: 1.2 &  33,108 &   723.9 $\pm$  2.6 &  -444.3 $\pm$  2.8 &   5.20 &   0.44 &  59.9 $\pm$  5.0 &  30.94 $\pm$ 9.86$^1$ \\
 5,228  & ...   &  7:16:21.9  &$ +23$:42:32.4 &  24,171 &   936.1 $\pm$  0.4 &  -571.1 $\pm$  0.4 &   6.30 &   1.30 &  23.6 $\pm$  1.1 &  17.8  $\pm$ 3.0$^3$  \\
 6,1923 & 3444  &  7:32:10.7  &$ +57$:55:33.7 &  16,145 &     2.7 $\pm$  0.6 &  -914.2 $\pm$  0.7 &   6.02 &   2.27 &  44.0 $\pm$  1.7 &  ~~~~~~~ ... ~~~~~~~~~      \\
 7,243  &  ...  &  8: 3: 7.1  &$ +34$:55:59.6 &  11,113 &  -512.7 $\pm$  0.7 & -1494.0 $\pm$  0.8 &   3.29 &   1.61 &  41.0 $\pm$  1.2 &   ~~~~~~~ ... ~~~~~~~~~    \\
 8,1976 & ...   &  8: 3: 7.1  &$ +52$:50:38.0 &  12, 33 &   362.3 $\pm$  2.3 &  -668.2 $\pm$  2.1 &   4.11 &   1.11 &  40.3 $\pm$  5.1 &  32.07 $\pm$  4.26$^1$\\ 
 9,2065 & 3517  &  8:53:46.0  &$ - 3$:29:36.0 &  11, 23 &  -507.1 $\pm$  1.5 &  -189.9 $\pm$  2.8 &   4.93 &   1.66 & 109.9 $\pm$  4.6 &  113.3 $\pm$ 2.0$^4$\\
10,306  & 3668  & 11:31:26.7  &$ -14$:58:17.0 &   8, 14 &   401.3 $\pm$  5.6 & -1378.2 $\pm$  5.2 &   6.02 &   1.78 &  85.0 $\pm$ 19.2 &  89.24 $\pm$ 1.69$^7$ \\
11,2472 & 452.1 & 11:54: 8.1 &$ + 9$:48:11.9 &   7, 36 &    90.0 $\pm$  1.5 &  -791.0 $\pm$  1.5 &   6.02 &   1.91 &  88.3 $\pm$  3.7 &    92 $\pm$  12$^5$   \\
12,326  & ...   & 12:24:23.2  &$ - 4$:43:47.2 &  13, 54 & -1142.3 $\pm$  1.7 &  -635.8 $\pm$  1.6 &   4.11 &   2.10 &  10.6 $\pm$  4.3 &   ~~~~~~~ ... ~~~~~~~~~           \\
13,334  & ...   & 12:34:14.5  &$ +20$:36:33.4 &   8, 34 &   333.7 $\pm$  1.3 & -1294.8 $\pm$  1.9 &   5.20 &   2.42 &  22.1 $\pm$  3.9 &   ~~~~~~~ ... ~~~~~~~~~   	\\
14,2632 & ...   & 12:46:50.8  &$ +31$:47:57.0 &  12,127 &  -796.6 $\pm$  0.8 &    40.9 $\pm$  0.6 &   6.30 &   2.05 &  53.3 $\pm$  2.3 &   ~~~~~~~ ... ~~~~~~~~~           \\
15,...  & 1167 A& 13: 9:32.1 &$ +28$:56:45.0 &   7, 14 &  -337.6 $\pm$  1.8 &  -207.9 $\pm$  2.9 &   3.83 &   1.88 &  92.5 $\pm$  5.0 &   84 $\pm$  12$^6$    \\
16,2686 & ...   & 13: 9:59.7  &$ +47$:44:40.5 &   5, 42 &  -634.2 $\pm$  1.2 &  -606.7 $\pm$  1.3 &   6.30 &   1.80 &  96.8 $\pm$  4.7 &   ~~~~~~~ ... ~~~~~~~~~   	\\
17,2719 & ...   & 13:20:32.2  &$ - 3$:55:23.0 &   5, 64 &  -850.4 $\pm$  0.9 &   287.5 $\pm$  1.3 &   6.57 &   2.00 &  37.1 $\pm$  3.4 &   ~~~~~~~ ... ~~~~~~~~~   	\\
18,360  & ...   & 13:46:53.6  &$ + 5$:43: 5.2 &  11, 33 &  -767.9 $\pm$  1.0 &  -848.0 $\pm$  1.3 &   4.38 &   2.05 &  17.8 $\pm$  3.0 &  10.33 $\pm$ 2.85$^4$ \\
19,369  & 545   & 14:20: 7.5 &$ - 9$:35:49.4 &   8, 28 &  -606.7 $\pm$  1.3 &  -829.2 $\pm$  2.1 &   5.20 &   1.85 &  69.0 $\pm$  4.0 &  71.52 $\pm$ 1.38$^7$ \\
20,2924 & 3849  & 14:28:36.2 &$ +33$:10:47.9 &  10, 29 &  -338.0 $\pm$  2.9 &  -714.8 $\pm$  2.5 &   4.38 &   1.15 &  84.5 $\pm$  5.5 &  90.0  $\pm$  1.3$^8$ \\
21,3343 & 4040  & 17:57:50.9 &$ +46$:34:53.3 &   9,155 &   -10.7 $\pm$  0.6 &   592.0 $\pm$  1.0 &   6.02 &   1.99 &  68.4 $\pm$  2.3 &  73.75 $\pm$  1.84$^1$\\ 
22,3445 & 9652 A& 19:14:37.5&$ +19$:18:31.2 &  23,153 &  -630.7 $\pm$  0.8 &   441.2 $\pm$  1.3 &   6.30 &   0.41 &  55.7 $\pm$  2.3 &  52.45 $\pm$  2.75$^1$\\ 
23,3482 & 767 A & 19:46:15.1 &$ +32$: 1: 0.8 &  74, 93 &   463.9 $\pm$  1.1 &  -393.8 $\pm$  1.7 &   6.02 &   0.28 &  68.6 $\pm$  3.4 &  74.90 $\pm$ 2.93$^1$ \\
24,3494 & 1245 A& 19:53:48.9&$ +44$:24:56.5 &  15,113 &   410.9 $\pm$  3.2 &  -469.9 $\pm$  1.5 &   4.11 &   0.52 & 227.9 $\pm$  3.9 &   219.2 $\pm$  1.4$^9$\\
25,528  & 1271  & 22:42:43.4 &$ +17$:40:12.1 &   9,104 &  1105.0 $\pm$  0.9 &   528.4 $\pm$  1.1 &   5.20 &   1.70 &  57.6 $\pm$  2.4 &  47.13 $\pm$  2.96$^1$\\ 
26,3872 & 4302  & 22:54:38.4 &$ - 5$:27:59.7 &   6, 32 &   609.9 $\pm$  1.5 &   354.3 $\pm$  2.0 &   5.20 &   2.19 &  40.6 $\pm$  4.6 &   ~~~~~~~ ... ~~~~~~~~~           \\
27,535  & 4312  & 23: 7:42.2 &$ +68$:41:51.5 &  40, 70 &  1139.0 $\pm$  1.3 &    60.3 $\pm$  1.4 &   5.48 &   0.78 &  76.7 $\pm$  2.8 &  63.5 $\pm$  4.2$^{10}$\\
\hline
\end{tabular}
\tablebib
{
  $^1$\citet{Perryman1997},        $^2$\citet{1944ApJ...100...55V},
  $^3$\citet{1985AJ.....90..123H}, $^4$\citet{2005AJ....130..337C},
  $^5$\citet{1959AJ.....64..269S}, $^6$\citet{1994AJ....108.2338H}, 
  $^7$\citet{2005AJ....129.1954J}, $^8$\citet{Monet1992}, 
  $^9$\citet{1993AJ....105.1571H}, $^{10}$\citet{1982AJ.....87..419D}
}
\tablefoot{$N_*$ = number of reference stars, $N_f$ = number of frames, $\Delta$T = epoch
range, COR = correction to absolute parallax.}
\end{table*}
%
%
In {Table} \ref{parallax} we report the TOPP results for the 27 red
dwarfs. These stars were originally selected for a number of reasons: 4 were
chosen to overlap with the USNO program, 4 were part of the TOPP
investigation into the catalogue of nearby stars \citep{Smart2007} and the rest 
are visual binary systems and high proper motions stars, historically
subjects of interest at the Torino Observatory.  In table \ref{parallax}
columns are: target names, positions, number of reference stars, number of
frames, proper motions, the interval of time for which we have observations,
correction applied to the relative parallax to obtain an absolute parallax,
absolute parallax estimates and literature values when available. Note the
proper motions are {essentially} relative, not absolute, hence any use of
these proper motions should be made with care.

To exhibit the variation in coverage in Figure \ref{twosolutions} we plot the
solutions of GJ 1167 A and LHS 1104 with{ 14 and 136 }observations respectively.
The reason for this large difference in observational history is a combination
of longer temporal coverage, higher priority (e.g. LHS 1104 as a visual binary
has a higher priority), and sporadic access to the telescope.

Measured parallax values of targets in common with \citet{Smart2007} (GJ 1167 A, LHS 2472, LHS 369, LHS 3872 and LHS
1050) agree within the errors
All objects agree with literature values to within
2$\sigma$ apart from LHS 1475, LHS 3494, LHS 528 and LHS 535 which are all
addressed in the discussion.

\section{Model Comparison}

\begin{table*}
\caption{\label{lumino} Photometric data and absolute magnitudes based on the parallaxes
    in {Table} \ref{parallax}.}
\centering
\begin{tabular}{llrrrrrrrr}
\hline\hline
ID &  LHS &    V~~      &    R~~      & I~~         & $J$  ~~  & $H$	~~  & $Ks$	~~  & M($J$)~~~~~~	  & M($K$)~~~~~~\\
\hline \\ 
1 &1047	& 11.48$^1$  &  10.27$^1$  &   8.71$^1$  & 7.22   & 6.71  & 6.39  & $8.49 ~  (-0.14,+0.14)        $ &$7.66 ~  (-0.14,+0.14)$ \\ 
2 &1050	& 12.61$^2$  &  11.46$^2$  &  10.04$^2$  & 8.62   & 8.07  & 7.81  & $8.33  ~ (-0.08, +0.09)$ &$7.52 ~ (-0.08, +0.09)$ \\ 
3 &1104	& 12.54$^3$  &  11.35$^3$  &  10.34$^3$  & 8.93   & 8.35  & 8.10  & $6.97  ~ (-0.12, +0.12)$ &$6.13 ~ (-0.12, +0.12)$ \\ 
4 &1475	& 10.48$^3$  &   9.50$^3$  &   8.71$^3$  & 7.43   & 6.80  & 6.59  & $6.31  ~ (-0.17, +0.19)$ &$5.48 ~ (-0.17, +0.19)$ \\ 
5 &228	& 15.53$^4$  &  14.42$^5$  &  13.25$^4$  & 12.02  & 11.52 & 11.3  & $8.89  ~ (-0.10, +0.11)$ &$8.17 ~ (-0.10, +0.11)$ \\ 
6 &1923	& 18.09$^4$  &  15.91$^5$  &  13.94$^4$  & 11.92  & 11.38 & 11.09 & $10.14 ~ (-0.08, +0.08)$ &$9.31 ~ (-0.08, +0.08)$ \\ 
7 &243	& 16.09$^6$  &  14.80$^6$  &  13.12$^6$  & 11.51  & 11.04 & 10.74 & $9.58  ~ (-0.06, +0.07)$ &$8.81 ~ (-0.06, +0.07)$ \\ 
8 &1976	& 11.38$^3$  &  10.31$^3$  &   9.41$^3$  & 8.06   & 7.48  & 7.24  & $6.08  ~ (-0.26, +0.29)$ &$5.26 ~ (-0.26, +0.29)$ \\ 
9 &2065	& 18.96$^7$  &  16.78$^7$  &  14.49$^7$  & 11.21  & 10.47 & 9.94  & $11.42 ~ (-0.09, +0.09)$ &$10.1 ~ (-0.09, +0.09)$ \\ 
10&306	& 14.19$^8$  &  12.80$^8$  &  11.05$^8$  & 9.36   & 8.76  & 8.50  & $9.00  ~ (-0.44, +0.56)$ &$8.14 ~ (-0.44, +0.56)$ \\ 
11&2472	& 12.81$^2$  &  11.61$^2$  &  10.12$^2$  & 8.70   & 8.19  & 7.87  & $8.43  ~ (-0.09, +0.09)$ &$7.6  ~ (-0.09, +0.09)$ \\ 
12&326	& 14.93$^9$  &  13.99$^9$  &  13.05$^9$  & 11.93  & 11.43 & 11.23 & $7.06  ~ (-0.74, +1.14)$ &$6.36 ~ (-0.74, +1.14)$ \\ 
13&334	& 18.02$^{10}$& 16.70$^{10}$ &15.13$^{10}$& 13.75  & 13.25 & 13.04 & $10.47 ~ (-0.35, +0.42)$ &$9.77 ~ (-0.35, +0.42)$ \\ 
14&2632	& 19.15$^4$  &  16.80$^5$  &  14.76$^4$  & 12.23  & 11.58 & 11.21 & $10.86 ~ (-0.09, +0.10)$ &$9.84 ~ (-0.09, +0.10)$ \\ 
15&...   & 14.52$^3$ &  13.35$^3$  &  11.88$^3$  & 9.48   & 8.91  & 8.61  & $9.31  ~ (-0.11, +0.12)$ &$8.44 ~ (-0.11, +0.12)$ \\ 
16&2686	& 14.16$^{11}$& 12.88$^{11}$&11.18$^{11}$ & 9.58   & 9.00  & 8.69  & $9.51  ~ (-0.10, +0.11)$ &$8.62 ~ (-0.10, +0.11)$ \\ 
17&2719	&  ...       &  14.84$^5$  &  13.09$^5$  & 11.28  & 10.77 & 10.48 & $9.12  ~ (-0.19, +0.21)$ &$8.33 ~ (-0.19, +0.21)$ \\ 
18&360	& 15.22$^9$  &  14.29$^9$  &  13.41$^9$  & 12.39  & 11.85 & 11.62 & $8.64  ~ (-0.34, +0.40)$ &$7.87 ~ (-0.34, +0.40)$ \\ 
19&369	& 12.96$^2$  &  11.75$^2$  &  10.25$^2$  & 8.74   & 8.19  & 7.98  & $7.93  ~ (-0.12, +0.13)$ &$7.17 ~ (-0.12, +0.13)$ \\ 
20&2924	& 19.58$^{12}$& 17.28$^5$  & 15.21$^{12}$ & 11.99  & 11.23 & 10.74 & $11.62 ~ (-0.14, +0.15)$ &$10.3 ~ (-0.14, +0.15)$ \\ 
21&3343	& 11.68$^3$  &  10.42$^3$  &   9.32$^3$  & 7.85   & 7.25  & 7.00  & $7.02  ~ (-0.07, +0.07)$ &$6.17 ~ (-0.07, +0.07)$ \\ 
22&3445	& 11.59$^3$  &  10.31$^3$  &   9.16$^3$  & 7.58   & 7.03  & 6.81  & $6.31  ~ (-0.09, +0.09)$ &$5.54 ~ (-0.09, +0.09)$ \\ 
23&3482	&  9.79$^3$  &   8.71$^3$  &   7.82$^3$  & 6.88   & 6.22  & 6.04  & $6.06  ~ (-0.11, +0.11)$ &$5.22 ~ (-0.11, +0.11)$ \\ 
24&3494	& 13.41$^3$  &  11.41$^3$  &   9.76$^3$  & 7.79   & 7.19  & 6.85  & $9.58  ~ (-0.04, +0.04)$ &$8.64 ~ (-0.04, +0.04)$ \\ 
25&528	& 11.76$^3$  &  10.56$^3$  &   9.52$^3$  & 8.06   & 7.38  & 7.18  & $6.86  ~ (-0.09, +0.09)$ &$5.98 ~ (-0.09, +0.09)$ \\ 
26&3872	& 13.87$^3$  &  12.43$^3$  &  11.20$^3$  & 9.65   & 9.09  & 8.81  & $7.69  ~ (-0.23, +0.26)$ &$6.85 ~ (-0.23, +0.26)$ \\ 
27&535	& 12.45$^3$  &  11.20$^3$  &  10.10$^3$  & 8.62   & 8.10  & 7.92  & $8.05  ~ (-0.08, +0.08)$ &$7.34 ~ (-0.08, +0.08)$ \\ 
\hline                                                                                                                                 
\end{tabular}
\tablebib
{    $^1$\citet{2007MNRAS.380.1261K}, 
    $^2$\citet{1990AAS...83..357B},  
    $^3$\citet{1996AJ....112.2300W}, 
    $^4$TOPP see text,
    $^5$GSC2.3 transformation see text, 
    $^6$\citet{1989PASP..101..614D}, 
    $^7$\citet{2005AJ....130..337C}, 
    $^8$\citet{2005AJ....129.1954J},
    $^9$\citet{2006AJ....132.1234C}, 
    $^{10}$\citet{2008AJ....136..840J},
    $^{11}$\citet{1979PASP...91..193W},
    $^{12}$\citet{2002AJ....124.1170D}}
\tablefoot{All infrared magnitudes come from the 2MASS catalogue. Absolute magnitude
  uncertainties include both parallax and apparent magnitude errors.}
\end{table*}
%
%
\subsection{Photometry}
We attempted to compile a set of uniform magnitudes in the $V, R_c, I_c, J, H,
K_s$ (hereafter the Cousins bands will be simply denoted $R$ and $I$) for the
comparison to theoretical models. The photometric data are from
various sources, where available we used photoelectric values from the
literature. As part of the TOPP program, many fields were observed in the $V$
and $I$ bands for the DCR correction and transformed to a standard system
using procedures described in \citet{2001AA...368..335B}. When an object had
no photometry in the literature we took the $V,I$ from the TOPP
observations. Overall, we obtained photoelectric photometry on the standard
$V,R,I$ systems for around 90\% of our objects. The errors for the
different sources varied from 0.02 to 0.08 magnitudes.

For the remaining missing magnitudes we used photographic magnitudes from the
GSC2.3 catalogue. The GSC2.3 system is described in \citet{LASKER2008} and
within the GSC-II consortium there are a set of standard transformations from
the Johnson-Cousins system to natural GSC-II plate bandpass. However, these
transformations apply to stars and the M dwarfs under study here
have extreme colors, so we constructed a transformation calibrated on an
identical sample. To do this we took a sample of M dwarfs from the RECONS
program and the Hipparcos catalogue to determine a linear transformation between
the GSC2.3 magnitudes, $V_{\rm pg}$, $R_{\rm F}$ and $I_{\rm N}$ and the
Johnson-Cousins magnitudes $V, R, I$ viz:
\begin{itemize}
\item[V] = V$_{\rm pg}$ + (0.566 $\pm$ 0.226) - ((0.384 $\pm$ 0.215)*(V$_{\rm pg}$ - R$_{\rm F}$))
\item[R] = R$_{\rm F}$ + (0.515 $\pm$ 0.022) - ((0.397 $\pm$ 0.013)*(R$_{\rm F}$ - I$_{\rm N}$))
\item[I] = I$_{\rm N}$ + (-0.006$\pm$ 0.033) - ((0.046 $\pm$ 0.011)*(R$_{\rm F}$ - I$_{\rm N}$))
\end{itemize}

The $R_{\rm F}$ and $I_{\rm N}$ filters transformations are well
determined while the $V_{\rm pg}$ filter is not. In table \ref{lumino} we
indicate those magnitudes obtained from this transformation with the index
$^5$.

All the infrared magnitudes come from the 2MASS catalogue. As these are in
general more precise than the optical measures and on a more consistent
system, we limit the use of the optical magnitudes to the color axis and when
possible use $R-K_s$ to give us a large baseline.

\begin{figure}[ht]
  \includegraphics[viewport=60 3 400 245,clip]{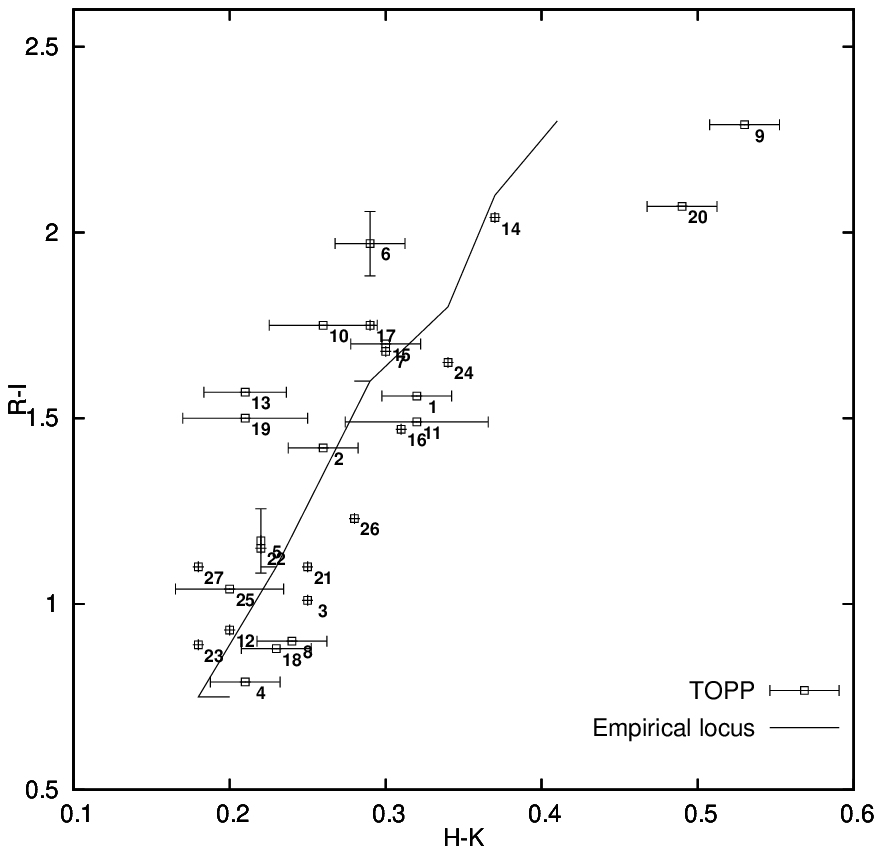} 
  \caption{$R-I$ versus $H-K_s$ diagram for the TOPP stars with identifiers as
    listed in Table \ref{parallax}. The continuous line is the mean observed
    colours from \citet{Leggett1992}.}
  \label{red}%
\end{figure}

In Figure \ref{red} plots of $R-I$ against $H-K_s$ have been made for TOPP stars
to compare with \citet{Leggett1992} mean observed colours for red dwarfs. The
$H-K$ values of \citet{Leggett1992} are in CIT system and they have been
transformed to 2MASS values using the following transformation from
\citet{Carpenter2001}.

$
  (K_s)_{\rm 2MASS} = K_{\rm CIT} + (0.000 \pm 0.005) (J-K)_{\rm CIT} - (0.024 \pm 0.003)\\
(H-K_s)_{\rm 2MASS} = (1.026 \pm 0.020) (H-K)_{\rm CIT} \ \ + \ \ (0.028 \pm 0.005)
$
%

\begin{figure}
  \includegraphics[viewport=65 2 400 245,clip]{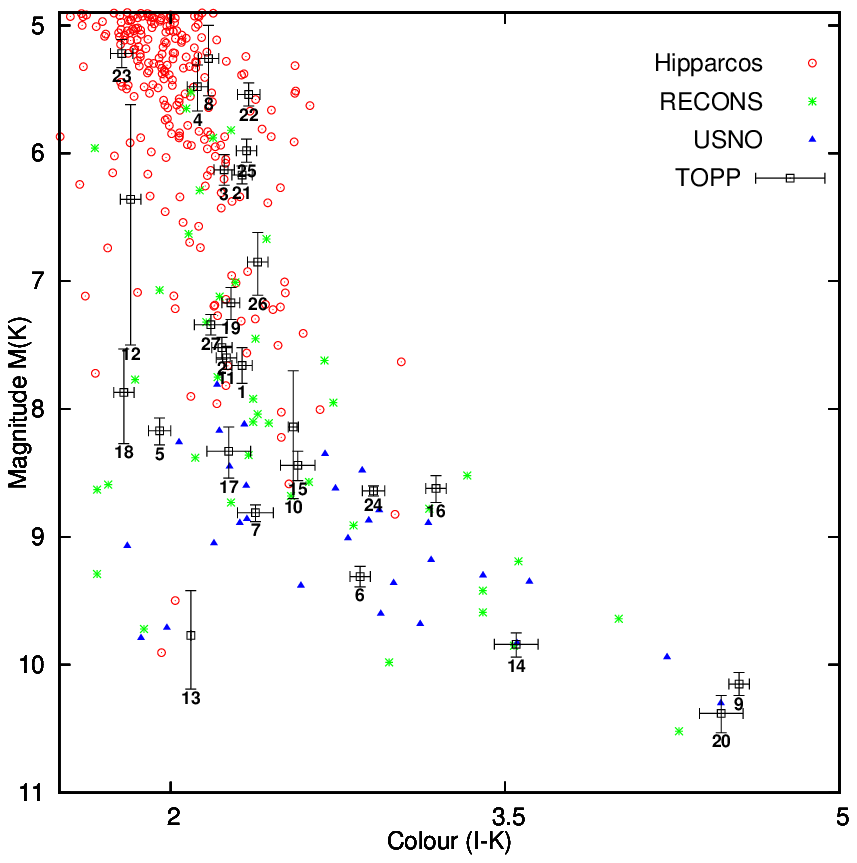} 
  \caption{Colour - Magnitude diagram ($M(K)$ vs $I-K_s$) of the TOPP stars
    identified as listed in Table \ref{parallax}, with Hipparcos, RECONS and
    USNO objects plotted for reference.}
  \label{hip}%
\end{figure}

As a comparison sample we selected 797 stars from the Hipparcos, 
RECONS and USNO catalogues with the following criteria:
$\pi/\sigma_{\pi} >$ 5, $\pi$ $>$ 20mas, and $V_J$ $>$ 6.  Figure \ref{hip}
shows the Colour - Magnitude diagram for TOPP stars and this reference
sample. The spread seen is partially due to errors but also intrinsic to the
sample.

\subsection{Theoretical models}
\begin{figure}
  \centering
  \includegraphics[viewport=65 0 400 252,clip]{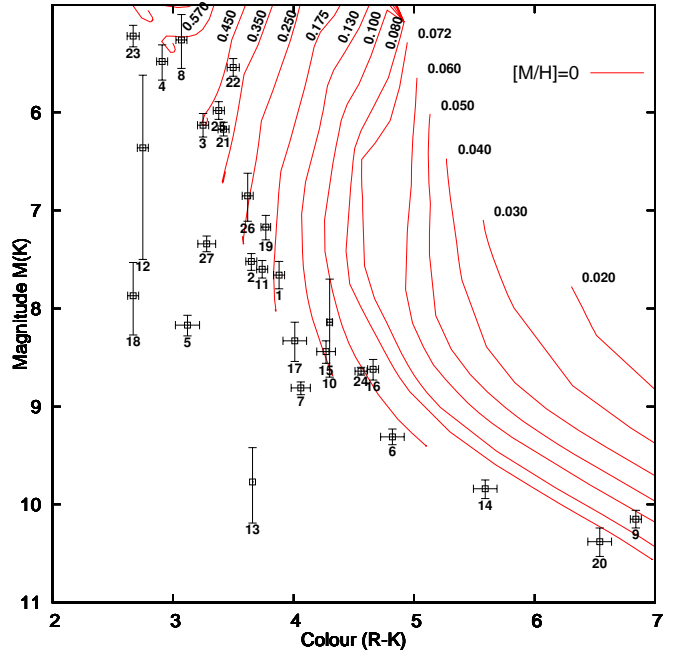}
  \includegraphics[viewport=65 0 400 252,clip]{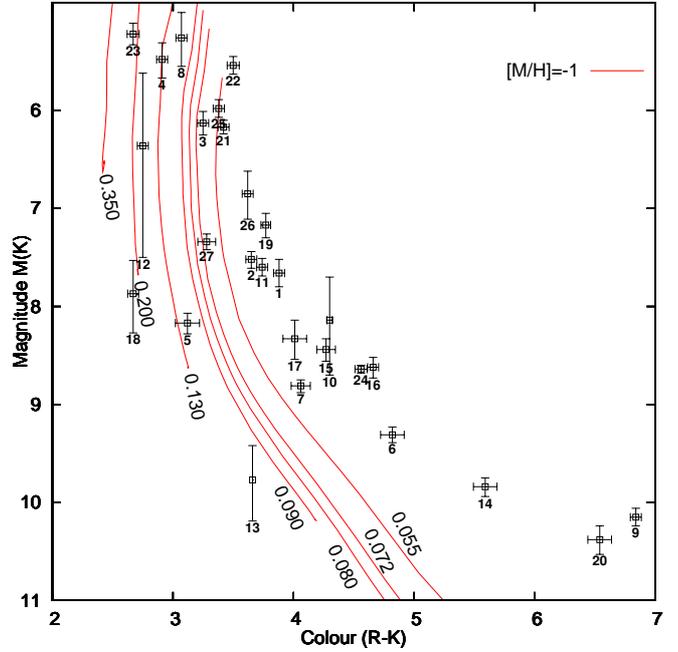}
  \caption{Colour - Magnitude diagram of the NextGen theoretical models for
    {[}M/H{]}=0, top panel, and {[}M/H{]}=-1.0, bottom panel. Lines represent
    different masses (labeled at the end of lines), points are TOPP results
    with identifiers as shown in Table \ref{parallax}}
  \label{MASS}%
\end{figure}

\begin{figure}
  \centering
  \includegraphics[viewport=65 0 400 252,clip]{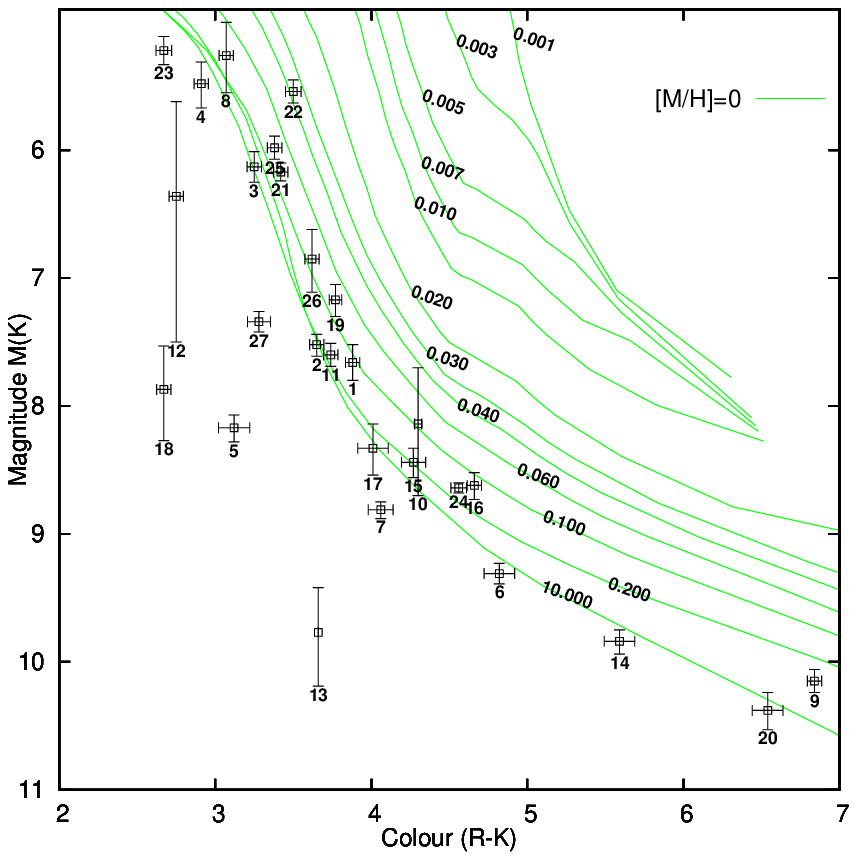}
  \includegraphics[viewport=65 0 400 252,clip]{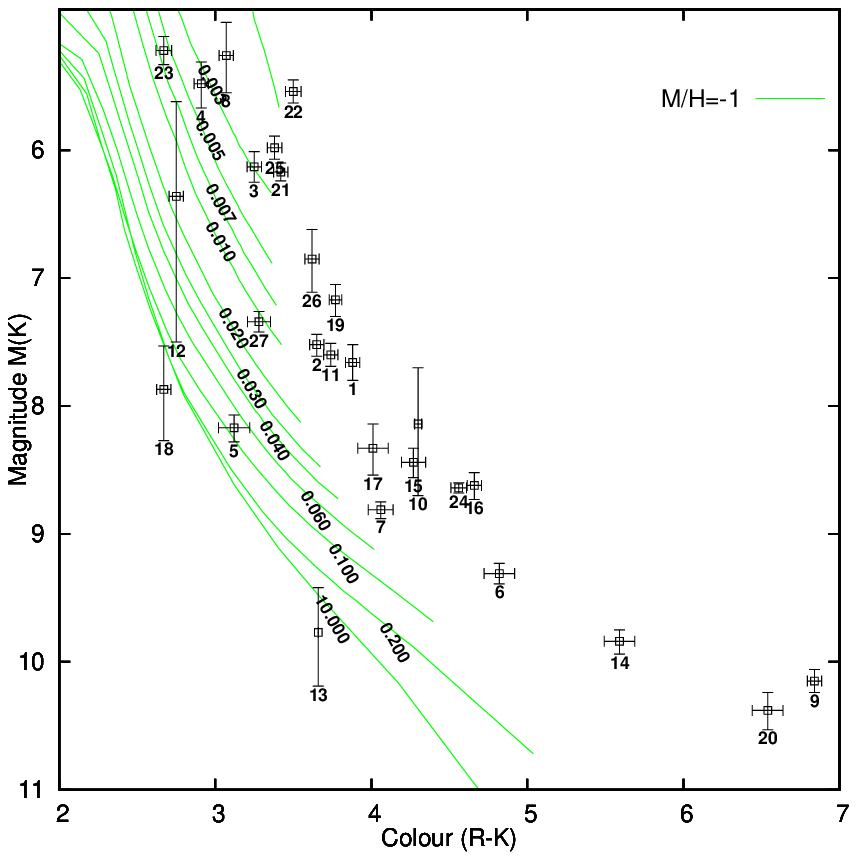}
  \caption{Colour - Magnitude diagram of the NextGen theoretical models for
    {[}M/H{]}=0, top panel, and {[}M/H{]}=-1.0, bottom panel. Lines represent
    different ages in Gyrs (labeled along the lines), points are TOPP results
    with identifiers as shown in Table \ref{parallax}}
  \label{AGE}%
\end{figure}

We compared a number of Color - Magnitude relations to theoretical model
simulations to find parameters for our targets. All simulations were generated
using the Phoenix web simulator ({http://phoenix.ens-lyon.fr/simulator}) with
the NextGen model \citep{1999ApJ...525..871H} for{ metallicities
  {[}M/H{]}} =0.0, -0.5, -1.0 and -1.5. We generated directly Cousins $RI$ and
2MASS $JHK_s$ magnitudes to avoid possible transform problems.  { As
  examples we plot TOPP results with low and high metallicity simulations with
  equal mass, Fig. \ref{MASS}, and age, Fig.  \ref{AGE}, contours.  }  It is
clear that different values of{ metallicity }are needed to describe our
stars. We did not feel the precision of the models or of the observations
warranted a finer grid in metallicity, which sometimes lead to conflicting
results, such as subdwarfs with a young age. Our procedure is to try to find
the closest to solar metallicity main sequence solution and then seek other
solutions if no solar metallicity solution can be found.  From the various
Colour - Magnitude diagram and theoretical model comparisons we estimate
masses and ages as shown in {Table} \ref{results}. In this table we also
include tangential velocities derived from the TOPP results and estimated
spectral types taken from the references listed. Finally, we check our
observational sequence for any possible variability and include this in the
discussion when relevant.

{ Metallicity assumptions influence the inferred masses and ages. For
  example the M3.5 LHS 1050 (object number 2 in Figures 5 and 6) is assigned
  [m/H]=0. If it is in fact more metal-poor with [m/H]=-0.5 then the assigned
  age would be younger (0.02-0.04 Gyr instead of 0.125-10 Gyr) and the mass
  would be lower (0.09-0.12 M$_\odot$ instead of 0.21-0.23
  M$_\odot$). Similarly, if the metallicity is higher, an older age and higher
  mass would be determined. For example if LHS 228 (object number 5) has a
  metallicity of -0.5 instead of -1, then the mass would be 0.13-0.175
  M$_\odot$ instead of 0.10-0.14 M$_\odot$ and the age would be $\>$ 10 Gyr
  instead of 0.05-0.20 Gyr. The effect is not a strong function of spectral
  type; for the M6 LHS 1923 (object number 6), if the metallicity is
  [m/H]=-0.5 instead of [m/H]=0, then again the assigned age would be very
  young (0.1-0.2 Gyr instead of 0.4-10 Gyr) and the mass would be lower (0.075
  M$_\odot$ instead of 0.1 M$_\odot$). Hence if the metallicity is unknown to
  0.5 dex, age is effectively undetermined and mass is uncertain to about a
  factor of two.  }

\section{Discussion}

\begin{table}
\caption{\label{results} Derived Parameters from Model Comparisons.}
\begin{tabular}{llllll}
\hline\hline
ID, LHS    & MK  & M/H & Mass        & Age         & V$_{tan}$     \\   
           & Type &       & (M$_\odot$)  & (Gyr)       &   (km/s) \\   
\hline   \\                                                                                                                     %
1 ,1047\tablefootmark{*}&M4$^1$     & 0   &  0.16-0.18    & 0.08-0.17   & 22  $\pm$ 1   \\ 
2 ,1050	 &  M3.5$^1$   & 0   &  0.21-0.23    & 0.125-10    & 37  $\pm$ 1   \\ 
3 ,1104	 &  	       & 0   &  0.42-0.46    & 0.12-10     & 59  $\pm$ 3   \\ 
4 ,1475	 &  M1$^1$     & 0   &  0.55-0.58    & 0.12-0.30   & 48  $\pm$ 4   \\ 
5 ,228	 &  	       & -1  &  0.10-0.14    & 0.05-0.20   & 206 $\pm$ 10  \\ 
6 ,1923	 &  M6$^2$     & 0   &  0.10-0.11    & 0.4-10      & 98  $\pm$ 4   \\ 
7 ,243	 &  	       & -.5 &  0.08-0.11    & 0.11-0.25   & 179 $\pm$ 5   \\ 
8 ,1976\tablefootmark{*}&	       & 0   &  0.53-0.61    & 0.055-10    & 83  $\pm$ 10  \\ 
9 ,2065	 &  M9e$^3$    & 0   &  0.06-0.07    & 0.25-0.45   & 23  $\pm$ 1   \\ 
10,306	 &  M5.5$^4$   & 0   &  0.10-0.13    & 0.04-10     & 80  $\pm$ 18  \\ 
11,2472	 &  M3.5$^1$   & 0   &  0.18-0.21    & 0.12-10     & 43  $\pm$ 2   \\ 
12,326	 &  	       & -.5 &  0.34-0.44    & 0.035-10    & 582 $\pm$ 238 \\ 
13,334	 &  	       & -1  &  0.09-0.10    & 0.3-10      & 286 $\pm$ 51  \\ 
14,2632	 &  M7.5$^5$   & 0   &  0.08-0.10    & 0.6-10      & 60  $\pm$ 3   \\ 
15,...	 &  M4$^1$     & 0   &  0.11-0.14    & 0.12-10     & 19  $\pm$ 1   \\ 
16,2686	 &             & 0   &  0.08-0.10    & 0.09-0.18   & 36  $\pm$ 2   \\ 
17,2719	 &             & 0   &  0.14-0.16    & 0.2-10      & 114 $\pm$ 11  \\ 
18,360	 &             & -1  &  0.20-0.35    & 0.2-10      & 304 $\pm$ 51  \\ 
19,369	 &  M4$^1$     & 0   &  0.18-0.22    & 0.060-0.105 & 70  $\pm$ 4   \\ 
20,2924	 &  M9e$^3$    & 0   &  0.07-0.08    & 0.6-10      & 43  $\pm$ 3   \\ 
21,3343	 &  M3$^6$     & 0   &  0.37-0.42    & 0.070-0.105 & 41  $\pm$ 1   \\ 
22,3445\tablefootmark{*}&M3$^1$     & 0   &  0.36-0.45    & 0.03-0.05   & 63  $\pm$ 3   \\ 
23,3482	 &  M1.5$^3$   & 0   &  0.60-0.62    & 0.2-10      & 38  $\pm$ 2   \\ 
24,3494\tablefootmark{*}&M5.5e$^3$  & 0   &  0.09-0.11    & 0.12-0.20   & 12  $\pm$ 0   \\ 
25,528	 &  M3$^1$     & 0   &  0.39-0.45    & 0.065-0.1   & 97  $\pm$ 4   \\ 
26,3872	 &  M4$^6$     & 0   &  0.23-0.28    & 0.06-0.125  & 82  $\pm$ 9   \\ 
27,535	 &  M3.5$^6$   & -.5 &  0.16-0.20    & 0.05-0.12   & 26  $\pm$ 1   \\ 
\hline 
\end{tabular}
\tablebib
{$^1$\citet{1985ApJS...59..197B}	
  $^2$\citet{1993ApJ...410..387F}	
  $^3$\citet{Leggett1992}
  $^4$\citet{1991AA...244..409M}
  $^5$\citet{1995AJ....109..797K}
  $^6$\citet{1991adc..rept.....G}}
\tablefoot{{ The mass and age ranges
    are based only on the errors in the parallax and photometry and do not
    include the systematic error of uncertain metallicity. The latter can have
    a large impact on mass and age as described in Section 4.2.}\\
  \tablefoottext{*}{These are known or suspect unresolved binary systems while the parameters
    are based on an assumption of being single stars. Unaccounted multiplicity
    will tend to increase the predicted age and decrease the predicted mass.}
}
\end{table}

{ We note that nearly half the sample in Table \ref{results} is
  predicted to be younger than 0.5 Gyr. There may be weak biases in our
  sample selection. The high proper motion selection could produce a bias to
  older objects. The choice of bright targets suitable for a 1m telescope may
  produce a bias to younger objects. However we do not believe this explains
  the apparently young ages in Table \ref{results}. A systematic underestimate
  of the metallicity (i.e. assigning a metallicity that is too low) could
  produce ages that are too young (see Section 4.2). This may suggest that a
  significant number of the nearby M dwarfs are metal-rich \citep[e.g. Figure
  2 of][]{2009ApJ...699..933J}, or it may reflect a systematic error in the
  modelled R and K magnitudes due to the known problems with the calculated
  opacities for cool atmospheres \citep{2007AA...473..257L}. }

{ The range of masses and ages in Table \ref{results} are estimated
  using only the uncertainties in the parallaxes and photometry, and do not
  include the systematic uncertainties due to model selection, in particular
  metallicity. As we show in Section 4.2 including this uncertainty implies
  that the age is not well determined and the masses could differ by a factor
  of two from what is listed in the table. We now consider groups of objects
  that appear to be similar.  }


\paragraph {\textit {\object{LHS 1050}, \object{LHS 1104}, \object{LHS 1923},
    \object{LHS 306}, \object{LHS 2472}, \object{GJ 1167 A}, \object{LHS
      2686}, \object{LHS 2719}, \object{LHS 369}, \object{LHS 3343},
    \object{LHS 3482}, \object{LHS 3872}, \object{LHS 528} }} are consistent
with being low-mass main-sequence stars with solar metallicities.  The LHS 528
parallax is over 2$\sigma$ from the Hipparcos value but the observational
sequence is solid and there is no evidence for binary motion, so we cannot
explain this inconsistency. However, we note that, based on the V magnitude of
11.76, this object is at the faint limit of the Hipparcos catalogue.

\paragraph {\textit {\object{LHS 1047}, \object{LHS 1976} }} are consistent
with being low-mass, single, main-sequence stars with solar metallicities,
however, the errors on the parallaxes are larger than average.  LHS 1047 was
discovered by \citet{1988AJ.....95.1226I} to be a binary system, speckle
measurements by \citet{2006AJ....132..994D} put the secondary 218 mas away
from the primary  {and 1.8 magnitudes fainter. }  LHS 1976 is seen in
the Hipparcos data as a binary system separated by 270 mas and speckle
interferometric observations \citep{2004AA...422..627B} also suggest the
presence of a third component 40 mas from the secondary  {and 0.27
  magnitudes fainter. If we assume the published magnitude differences the
  derived primary age range will be increased from (0.08-0.17) to (0.1-0.2)
  Gyr for LHS 1047 and (0.055-0.1) to (0.2-10) Gyr for LHS 1976; the mass
  range will not vary for LHS 1047 while for LHS 1976 it will decrease from
  (0.53-0.61) to (0.48-0.56) M$_\odot$. This does not take into account the
  effect of the unmodelled multiplicity on our parallax determination which is
  also probably the cause of the large parallax errors.}

\paragraph{\textit {\object{LHS 2065}, \object{LHS 2632}, \object{LHS 2924}}}
are objects close to the end of the main sequence with very low masses
(0.06-0.095), indicating that they are around the transition between low-mass
stars and brown dwarfs. All masses, ages and errors are based on assuming
solar metallicity.  LHS 2065 and LHS 2924 are catalogued as flare stars but the
only evidence of variability over the 4 years of TOPP observations is a
gradual dimming for LHS 2065 by 0.06+/-0.02 magnitudes.  LHS 2065 and LHS 2924
are also the reddest objects in our study and in fact they are outliers in
Figure \ref{red}; this can be most readily explained by the presence of dust
in their atmospheres \citep*[e.g.][]{Jones1997}. The lack of lithium in LHS
2065 \citep{martin99} indicates that the mass of this object is greater than
0.06M$_\odot$.

\paragraph {\textit {\object{LHS 1475}}} is a main sequence star in a visual
binary system with LHS 1476. We note that the TOPP parallax is twice the
Hipparcos value; we believe this is because during the Hipparcos mission this
object moved in front of another bright star and was effected by
veiling glare \citep{1985ashc.rept..133F}. This is also reflected in the large
errors of the Hipparcos values.  The Yale Parallax Catalog lists five
published individual parallaxes: 76.9 $\pm$ 17.8 and 60 $\pm$ 15.0 from the
M$^c$Cormick Observatory, 43.2 $\pm$ 5.0 and 36.9 $\pm$ 5.7 from the Sproul
Observatory and 46.3 $\pm$ 21.6 from the Pulkovo Observatory. We note that the
TOPP value is more consistent with these values than the Hipparcos value but
we postpone any further conclusions until a binary solution is made including
LHS 1476.

%
%

\paragraph {\textit {\object{LHS 3494}}} was cataloged as a double system by
\citet{1988ApJ...333..943M}  {with a secondary that is 1.1 magnitudes fainter.
If we assume this magnitude differences the derived primary age range will be
increased from (0.12-0.2) to ( 0.2-10.) Gyr and the mass range decreased from
(0.09-0.11) to (0.11-0.12) M$_\odot$.}  It is also a known flare star and
indeed the photometric standard deviation for the TOPP sequence was
approximately twice that of the other objects in this study.  The TOPP
parallax is inconsistent at the 2$\sigma$ level with that found by
\citet{1993AJ....105.1571H}, which may be due to the unmodelled binary nature.

\paragraph {\textit {\object{LHS 3445}}} The position of this object suggests
either a young star or a close binary system. Since this is part of a visual
binary system with LHS 3446 and we have used a simple single star solution it
is not possible to see if the fit residuals indicate an unresolved binary
companion.  \citet{2009ApJ...699..649S} find an age indicator of 60-300 Myr
which is inconsistent with our age (30-50 Myr) assuming the object is
single. Taking the radial velocity of -42.3 km/s from
\citet{2002AJ....123.3356G} combined with the TOPP results and the velocity of
the Sun from \citet{1998MNRAS.298..387D} we find a space motion, in km/s, of
U=$-25$ (radially inwards), V= $-26$ (in the direction of Galactic rotation),
and W=$69$ (vertically upwards). This velocity is not consistent with other
younger population velocities \citep[e.g. see][]{2009AA...501..941H,
  2004ARAA..42..685Z}. Finally, the photometric distance based on the color
and apparent magnitude and following \citet{Henry2004} is 12pc which, being
less than the trigonometric distance of 18pc, is consistent with a binary
system. Our results therefore imply that this object is more probably a binary
rather than a very young star.  {The effect of unaccounted multiplicity
  will be to increase the predicted age and decrease the predicted mass.}


\paragraph {\textit {\object{LHS 243}, \object{LHS 326}}} are consistent with
{[}M/H{]} = -0.5 models, hence subdwarfs. The parallax error for LHS 326 is
very high but still 2$\sigma$ away from the solar metallicity locus. { The
position of LHS 243 in Fig. 5 is most consistent with [M/H]=-0.5; however, if
it is closer to [M/H]=0 then the age range will increase from 0.11-0.25 to
0.25-10.0 Gyr, which is more consistent with our age expectation for a slightly
metal poor subdwarf.} The high tangential velocities of both objects suggest a
non-thin disk membership, supporting this classification.

\paragraph {\textit {\object{LHS 535}}} appears to be most consistent with
{[}M/H{]} = -0.5; however, this requires also a very young age and a
closer-to-solar metallicity would relax that requirement. We do not believe
that the data or models support an investigation using a finer grid, so
resolution of this inconsistency will have to wait until better models or
spectra are available. We note that the parallax is over 2$\sigma$ from the
USNO value but the observational sequence is solid and there is no evidence
for binary motion so we cannot explain the high difference.


\paragraph {\textit {\object{LHS 228}, \object{LHS 334},\object{LHS 360}}}
The comparison to models suggests that these object are subdwarfs and we base
the mass and age values on a {[}M/H{]} = -1 model. This interpretation is
supported by the high tangential velocities. LHS 228 also has a high radial
velocity \citep[73$\pm$3 km/s]{1989AJ.....98.1472D}, and a spectroscopic
classification of subdwarf \citep{2007ApJ...669.1235L}. { The
  metallicity of LHS 228 could however be between -1.0 and -0.5; this would
  increase the age range from 0.11-0.25 to 0.25-10.0 Gyr making it more
  consistent with the expectation for a subdwarf.  LHS 360 would fit a lower
  metallicity model but that requires a very young age.}

\section{Conclusions}
We have measured the parallaxes of 27 cool dwarf stars and compared them to a
set of models.  A number of high proper motion objects, planetary nebulae and
cataclysmic variables observed in the TOPP remained unpublished;
we plan to release these in the near future.  There are no plans to begin any
more large programs with the Torino telescope in the near future. If the Gaia
mission, in which Torino is heavily involved, is successful then there will no
longer be any need for parallaxes from 1m-class telescopes.

The majority of ground-based parallax programs are now concentrated on the
newly discovered brown dwarfs. These have very red colours and will be beyond
the magnitude limit of Gaia, requiring very red or infrared observations on
4m class telescopes. The denser and more precise reference frame from Gaia
will enable us to improve our reduction strategy and the TOPP observations 
remain a development test bed for this. Current activity is focused on the
calculation of the relative-to-absolute parallax correction, treatment of
binaries, and reference frame selection.

%
%
%
%
%

\section{Acknowledgments}
We would like to thank the referee, Dr S. K. Leggett, who's comments significantly
improved the quality of this paper.

HRAJ and RLS acknowledge the support of Royal Society International Joint
Project 2007/R3. BB, MGL and RLS acknowledge the support of INAF through the
PRIN 2007 grant n. CRA 1.06.10.04.

This research has made use of: the SIMBAD database operated at CDS France, the
Second Guide Star Catalog developed as a collaboration between the Space
Telescope Science Institute and the Osservatorio Astronomico di Torino, and,
the Two Micron All Sky Survey which is a joint project of the University of
Massachusetts and the Infrared Processing and Analysis Center/California
Institute of Technology.

\bibliographystyle{aa} 
\bibliography{local.bib}

\end{document}